\documentclass[12pt,reqno]{amsart}

\usepackage{amsmath,amsfonts,amsthm,cite,amssymb}

\usepackage{fullpage}
                                                                                
\usepackage{txfonts}

\newcommand{\I}{\mathrm{i}}
\newcommand{\E}{\mathrm{e}}

\usepackage{titlesec}
\titleformat{\section}
{\bfseries\scshape\centering}
{\thesection.}{.5em}{}
\titleformat{\subsection}
{\rmfamily\bfseries}
{\thesubsection}{.5em}{}
\newtheorem{theorem}{Theorem}

\begin{document}


\renewcommand{\thefootnote}{\fnsymbol{footnote}}
\baselineskip 16pt
\parskip 8pt
\sloppy




\title{Transformation Formula of the ``2nd'' Order Mock Theta Function}


\author{Kazuhiro \textsc{Hikami}}


  \address{Department of Physics, Graduate School of Science,
    University of Tokyo,
    Hongo 7--3--1, Bunkyo, Tokyo 113--0033,   Japan.
    }
    
    \urladdr{http://gogh.phys.s.u-tokyo.ac.jp/{\textasciitilde}hikami/}

    \email{\texttt{hikami@phys.s.u-tokyo.ac.jp}}


\date{September 21, 2005. Revised on October 24, 2005}



\begin{abstract}
We give a transformation formula for the ``2nd order'' mock theta
function
\begin{equation*}
  D_5(q)=
  \sum_{n=0}^\infty
  \frac{
    (-q)_n
  }{
    (q; q^2)_{n+1}
  } \, q^n
\end{equation*}
which was recently proposed in connection with the quantum invariant
for the Seifert manifold.

\end{abstract}





\maketitle

\section{Introduction and Statement of Results}

We study the transformation formula for the $q$-series $D_5(q)$
defined by
\begin{align}
  \label{define_D}
  D_5(q)
  & =
  \sum_{n=0}^\infty
  \frac{(-q)_n}{
    (q; q^2)_{n+1}
  } \,
  q^n 
  \\
  & =
  1 + 2 \, q + 4 \, q^2 + 6 \, q^3 + 10 \, q^4 + 16 \, q^5 + 23 \, q^6
  + \cdots
  \nonumber
\end{align}
where as usual we mean $q=\exp(2 \, \pi \, \I \, \tau)$ with
$\tau \in \mathbb{H}$.
This $q$-series can be rewritten as
\begin{equation}
  D_5(q)
  =
  \frac{1}{
    \left[ (q;q^2)_\infty \right]^2
  } \,
  \sum_{n=0}^\infty
  \left[ (q; q^2)_n \right]^2 \, q^{2n}
\end{equation}
which can be proved by setting
$\alpha=q$, $\beta=q$, $\gamma=0$, $z=q^2$
in the following
transformation formula of the $q$-hypergeometric functions
(see, \emph{e.g.}, Ref.~\citen{GEAndre66a})
\begin{equation}
  \sum_{n=0}^\infty
  \frac{
    (\alpha ;q^2)_n \,
    (\beta)_{2n}
  }{
    (q^2; q^2)_n \,
    (\gamma)_{2n}
  } \,
  z^n
  =
  \frac{
    (\beta)_\infty \, (\alpha \, z ; q^2)_\infty
  }{
    (\gamma)_\infty \, (z ; q^2)_\infty
  } \,
  \sum_{m=0}^\infty
  \frac{
    \left( \frac{\gamma}{\beta} \right)_m \,
    (z ; q^2)_m
  }{
    (q)_m \, (\alpha \, z ; q^2)_m
  } \,
  \beta^m  .
\end{equation}

The $q$-series $D_5(q)$
was introduced in Ref.~\citen{KHikami05b} in connection with the
quantum invariant of 3-manifold.
The definition~\eqref{define_D} can be extended to $|q|>1$, and we can
introduce a new $q$-series by replacing $q$ with $1/q$
in the summand of~\eqref{define_D};
\begin{align}
  D_5^*(q)
  & =
  \sum_{n=0}^\infty (-1)^n \, q^{n (n+1)/2} \,
  \frac{
    (-q)_n
  }{
    (q;q^2)_{n+1}
  }
  \\
  & =
  1 - q^2 + q^6 - q^{12} + q^{20} - q^{30} + q^{42} + \cdots
  \nonumber
\end{align}
A limiting value of $D_5^*(q)$ when $q$ is the root of unity  is
related to the SU(2) Witten--Reshetikhin--Turaev (WRT) invariant
$\tau_N(\mathcal{M})$~\cite{EWitt89a,ResheTurae91a} for 3-manifold
$\mathcal{M}$.
Precisely, we have~\cite{KHikami05a,KHikami05b}
\begin{equation}
  \label{WRT_222}
  \left(
    \E^{2 \pi \I/N} - 1
  \right) \cdot
  \tau_N( M(2,2,2)  )
  =
  2 \, 
  \left(
    1 - 2 \, D_5^*(\E^{\pi \I/N})
  \right)
\end{equation}
Here
$M(2,2,2)$ denotes the Seifert prism  manifold
$Oo0(-1; (2,1), (2,1), (2,1))$
(see, \emph{e.g.}, Refs.~\citen{JMiln75a,Montesi87Book}).

Lawrence and Zagier pointed out~\cite{LawrZagi99a}
that the Eichler integral of the modular form with weight $3/2$ has a
\emph{nearly} modular property,
and that  the WRT invariant for the
Poincar{\'e} homology sphere is regarded as a limiting value of the
Eichler integral.
This result is further extended to the WRT invariant for other
Seifert fibered manifolds, and
especially 
it was shown that the WRT
invariant~\eqref{WRT_222} for the prism manifold $M(2,2,2)$ has  a
nearly modular
property
under a transformation $1/N \leftrightarrow
-N$~\cite{KHikami05a}.

Well known is that the Ramanujan mock theta functions
satisfy this type of transformation
formula~\cite{Ramanujan87Book,GWats36}.
It  is demonstrated in
Refs.~\citen{LawrZagi99a,KHikami05b}
that the Eichler integral which reduces to  the WRT invariant for
the Poincar{\'e} homology sphere  gives the 5th order Ramanujan  mock theta
function
when we replace $q$ by $1/q$.
Therefore  the $q$-series $D_5(q)$ defined in~\eqref{define_D} is expected
to be a mock theta function \emph{\`{a} la} Ramanujan.
In this short note, we shall give the transformation formula for the
function $D_5(q)$, and we  prove that the function is indeed the mock
theta function.

Among Ramanujan's mock theta functions,
we recall 
the third order mock theta functions~\cite{GWats36} defined by
\begin{align}
  \label{define_omega}
  \omega(q)
  & =
  \sum_{n=0}^\infty
  \frac{
    q^{2 n (n+1)}
  }{
    \left[
      (q; q^2)_{n+1}
    \right]^2
  }
  \\
  & =
  \sum_{n=0}^\infty
  \frac{q^n}{(q;q^2)_{n+1}}
  \\[2mm]
  f(q)
  & =
  \sum_{n=0}^\infty
  \frac{
    q^{n^2}
  }{
    \left[
      (-q)_n
    \right]^2
  }
  \\
  & =
  2- \sum_{n=0}^\infty
  (-1)^n
  \, \frac{q^n}{(-q)_n}
\end{align}
Here the second equalities are proved in Ref.~\citen{NJFine88Book}.
As was proved in Ref.~\citen{GWats36}, these functions can be written
in the form  of the Lerche sum;
\begin{gather}
  (q^2; q^2)_\infty \cdot \omega(q)
  =
  \sum_{n \in \mathbb{Z}} (-1)^n \,
  \frac{
    q^{3n(n+1)}
  }{
    1 - q^{2n+1}
  }
  \\[2mm]
  (q)_\infty \cdot f(q)
  =
  2 \sum_{n \in \mathbb{Z}}
  (-1)^n \,
  \frac{
    q^{\frac{1}{2} n (3n+1)}
  }{
    1 + q^n
  }
\end{gather}

To study  the transformation formula for $D_5(q)$, we introduce two
functions;
\begin{gather}
  \begin{aligned}
    \label{define_H1}
    h_1(q)
    & =
    \sum_{n=0}^\infty
    \frac{
      (-q)_{2 n}
    }{
      \left[
        (q;q^2)_{n+1}
      \right]^2
    } \,
    q^n
    \\
    & =
    1+ 3 \, q + 7 \, q^2 + 14 \, q^3 + 27 \, q^4 + 49 \, q^5 + 84 \,
    q^6 + \cdots
  \end{aligned}
  \\[2mm]
  \begin{aligned}
    \label{define_H2}
    h_2(q)
    & =
    \sum_{n=0}^\infty
    (-1)^n \,
    \frac{
      (q; q^2)_n
    }{
      \left[
        (-q^2;q^2)_n
      \right]^2
    } \,
    q^{n^2}
    \\
    & =
    1 - q + q^2 + 2\, q^3 - q^4 - 4 \, q^5 + q^6 + \cdots
  \end{aligned}
\end{gather}
As was shown in Ref.~\citen{AndreBernd05a} as
Entry~12.3.9 and Entry~12.2.1
which originally appeared in Ref.~\citen{Ramanujan87Book},
those functions  can also be rewritten as the form of the Lerche sum
as follows;
\begin{gather}
  \label{define_h1}
  h_1(q)
  =
  \frac{1}{2} \,
  \frac{
    (-q)_\infty
  }{
    (q)_\infty
  } \,
  \sum_{n \in \mathbb{Z}}
  (-1)^n \,
  \frac{q^{n(n+2)}}{ 1 -q^{2n+1}}
  \\[2mm]
  \label{define_h2}
  h_2(q)
  =
  \frac{
    (q)_\infty
  }{
    \left[
      (q^2; q^2)_\infty
    \right]^2
  } \,
  \sum_{n \in \mathbb{Z}}
  \frac{
    q^{\frac{1}{2} n(n+1)}
  }{
    1+q^n
  }
  +
  \frac{1}{2} \,
  \frac{
    \left[
      (q)_\infty
    \right]^5
  }{
    \left[
      (q^2 ; q^2)_\infty
    \right]^4
  }
\end{gather}

As Entry~12.4.5 in Ref.~\citen{AndreBernd05a} 
proves
\begin{equation}
  \label{D_as_sum}
  D_5(q)
  =
  2 \, h_1(q)
  -
  \left[
    \frac{
      (q^2;q^2)_\infty
    }{
      (q)_\infty
    } 
  \right]^2 \,
  \omega(q)
\end{equation}
we need to study the transformation formulae for the functions
$h_1(q)$ and  the  third order mock theta function $\omega(q)$
defined in~\eqref{define_omega}.

In Ref.~\citen{GWats36}
Watson derived the following transformation of the third order mock
theta functions, $\omega(q)$ and $f(q)$;
\begin{equation}
  \label{Watson_transform}
  \int_0^\infty
  \E^{-\frac{3}{4} \alpha x^2} \,
  \frac{\cosh( \alpha \, x/2)}{
    \cosh(3 \, \alpha \, x/2)} \,
  \mathrm{d} x
  =
  - \sqrt{\frac{4 \, \pi}{3 \, \alpha}} \,
  q^{2/3} \,
  \omega(q)
  +
  \frac{1}{\sqrt{3}} \, \frac{\pi}{\alpha}
  \, q_1^{~-1/12} \,
  f(q_1^{~2})
\end{equation}
where  we have set
\begin{align}
  \label{define_q_alpha}
  q &=\E^{-\alpha}
  &
  q_1 & = \E^{-\pi^2/\alpha}
\end{align}
with $\Re \alpha>0$.

As will be proved in Section~\ref{sec:transformation}, we have
the following transformation formula for our functions $h_1(q)$ and $h_2(q)$;
\begin{theorem}
  \label{thm:main}
  Let the functions $h_1(q)$ and $h_2(q)$ be defined
  in~\eqref{define_h1} and~\eqref{define_h2} respectively.
  We have
  \begin{equation}
    \label{H_transform}
    \int_{-\infty}^\infty
    \frac{\E^{- \alpha x^2}}{
      \cosh( \alpha \, x)
    } \, \mathrm{d} x
    =
    -
    4  \, \sqrt{\frac{\pi}{\alpha}} \,  q^{3/4} \,
    h_1(q)
    +
    \frac{\pi}{\alpha} \,
    q_1^{~-1/4} \,
    \left(
      h_2(q_1^{~2})
      -
      \frac{1}{2} \,
      \frac{
        \left[ 
          (q_1^{~2} ; q_1^{~2})_\infty
        \right]^5
      }{
        \left[
          (q_1^{~4} ; q_1^{~4})_\infty
        \right]^4
      }
    \right)
  \end{equation}
  where we have set parameters $q$ and $q_1$ as
  in~\eqref{define_q_alpha}
  with  $\Re \alpha> 0$.
\end{theorem}

The left hand side of \eqref{H_transform} is the Mordell integral, and  the properties
of this integral was studied in detail in Ref.~\citen{LJMorde33a}
(see also Refs.~\citen{GEAndre81e,RamanujanCollected}).

This theorem proves that
the function $D_5(q)$ can be written as a sum of two mock theta
functions,
$h_1(q)$ and $\omega(q)$,
as in~\eqref{D_as_sum}.
Thus  we can conclude
that the $q$-series $D_5(q)$, which was originally introduced based on
studies of the WRT invariant,
is indeed
the mock theta function,
and that the transformation formula of $D_5(q)$ is given by combining 
\eqref{D_as_sum},
\eqref{Watson_transform},
and
\eqref{H_transform}.

\section{Proof of Theorem~\ref{thm:main}}
\label{sec:transformation}

Both the
integral in the left hand side
and the $q$-series in the right hand side
(see definitions~\eqref{define_H1} and~\eqref{define_H2})
of~\eqref{H_transform}
are analytic
in $\Re \alpha>0$.
Accordingly
we can assume
$\alpha \in \mathbb{R}_{>0}$
to simplify our proof
thanks to the analytic continuation.

We study  the integral defined by
\begin{equation}
  \label{define_K}
  K=
  \frac{1}{2 \, \pi \, \I}
  \int\limits_{\mathcal{C}}
  \frac{
    \pi}{
    \sin(\pi \, z)
  } \,
  \frac{
    \E^{- \alpha z (z+2)}
  }{
    1 - \E^{- \alpha (2z+1)}
  } \,
  \mathrm{d} z
\end{equation}
where a contour $\mathcal{C}$ encircles a real axis counterclockwise.
As the integrand converges to zero in $\Re z\to\pm\infty$,
the integral $K$ can be decomposed
into
\begin{equation*}
  K=
  \frac{1}{2 \, \pi \, \I}
   \left(
     \int_{-\infty - \I 0}^{\infty - \I 0}
     +
     \int_{\infty+ \I 0}^{-\infty+ \I 0}
   \right)
  \frac{
    \pi}{
    \sin(\pi \, z)
  } \,
  \frac{
    \E^{- \alpha z (z+2)}
  }{
    1 - \E^{- \alpha (2z+1)}
  } \,
  \mathrm{d} z
\end{equation*}
which we set $K= K_1 + K_2$.
It is easy to see that,
by the Cauchy theorem,
the integral $K$~\eqref{define_K} is  given by
\begin{equation}
  K =
  \sum_{n \in \mathbb{Z}} (-1)^n \, \frac{q^{n (n+2)}}{1- q^{2n+1}}
  -
  \frac{\pi}{2 \, \alpha} \, q^{-3/4}
\end{equation}

We first consider the integral $K_2$.
On the path contour of $K_2$ which is in the upper half plane
we can apply the
Fourier expansion
\begin{equation}
  \frac{1}{\sin (\pi \, z)}
  =
  -2 \, \I \,
  \sum_{n=0}^\infty
  \E^{(2n+1) \pi \I z}
\end{equation}
and substituting this expression, we obtain
\begin{equation}
  K_2
  =
  \frac{1}{2 \, \pi \, \I}
  \sum_{n=0}^\infty
  \int_{- \infty + \I 0}^{\infty + \I 0}
  F_n(z) \,
  \mathrm{d} z
\end{equation}
where the integrand $F_n(z)$ is defined by
\begin{equation}
  F_n(z)
  =
  2 \, \pi \, \I \,
  \frac{
    \E^{(2n+1) \pi \I z - \alpha z^2 - 2 \alpha z}
  }{
    1 - \E^{- \alpha (2 z+1)}
  }
\end{equation}
We see that the integrand $F_n(z)$ has simple poles at
\begin{equation*}
  z =\frac{m}{\alpha} \, \pi \, \I - \frac{1}{2}
  \equiv z_m
\end{equation*}
for $m \in \mathbb{Z}$, and the residues of $F_n(z)$ at $z=z_m$
are computed to be
\begin{equation}
  \lambda_{n,m}
  =
  \frac{\pi \, \I}{\alpha} \,
  \E^{(2n+1) z_m \pi \I - 
    \alpha z_m ( z_m + 2)
  }
\end{equation}
Then we obtain
\begin{align*}
  \frac{1}{ 2 \, \pi \, \I}
  &
  \sum_{n=0}^\infty
  \left(
    \int_{- \infty + \I 0}^{\infty + \I 0}
    -
    \int_{- \infty + \frac{2n+1}{2 \alpha} \pi \I }^{ \infty +
      \frac{2n+1}{2 \alpha} \pi \I }
  \right)
  F_n(z) \,
  \mathrm{d} z
  \\
  & =
  \sum_{n=1}^\infty
  \left(
    \lambda_{n,1}  +     \lambda_{n,2} 
    + \cdots +
    \lambda_{n,n} 
  \right)
  \\
  & =
  \sum_{m=1}^\infty
  \left(
    \lambda_{m,m} + \lambda_{m+1,m} + \cdots
  \right)
  \\
  & =
  \sum_{m=1}^\infty \lambda_{m,m} \,
  \frac{1}{
    1 - \E^{2 \pi \I z_m}
  }
  \\
  & =
  \frac{\pi}{\alpha} \,
  \E^{\frac{3}{4} \alpha} \,
  \sum_{m=1}^\infty
  \frac{
    q_1^{~m(m+1)}
  }{
    1 + q_1^{~2 m}
  }
\end{align*}
On the other hand, we have
\begin{align}
  \frac{1}{ 2 \, \pi \, \I}
  \int_{- \infty + \frac{2n+1}{2 \alpha} \pi \I }^{ \infty +
    \frac{2n+1}{2 \alpha} \pi \I }
  F_n(z) \,
  \mathrm{d} z
  & =
  \int_{-\infty}^{\infty}
  F_n\left(
    x + \frac{2 \, n+1}{2 \, \alpha} \, \pi \, \I
    - \frac{1}{2}
  \right)\,
  \mathrm{d} x
  \nonumber
  \\
  & =
  -  \pi \, \I \,
  q^{-3/4} \,
  q_1^{~ (2n+1)^2/4} \,
  \int_{-\infty}^\infty
  \frac{\E^{-\alpha x^2}}{
    \cosh(\alpha \, x)
  } \,
  \mathrm{d} x
\end{align}
Combining these results, we obtain
\begin{align}
  K_2 & =
  \frac{\pi}{\alpha} \,
  q^{-3/4} \,
  \sum_{m=1}^\infty
  \frac{
    q_1^{~m(m+1)}
  }{
    1 + q_1^{~2 m}
  }
  -
  \frac{1}{2}
  q^{-3/4} \,
  \left(
    \sum_{n=0}^\infty
    q_1^{~(2n+1)^2/4}
  \right) \,
  \int_{-\infty}^\infty
  \frac{
    \E^{- \alpha x^2}
  }{
    \cosh(\alpha \, x)
  } \,
  \mathrm{d} x
\end{align}

Integral $K_1$  from the lower half plane can be computed in the same manner.
Applying the Jacobi triple product formula, 
we finally obtain the transformation formula~\eqref{H_transform}
which proves 
Theorem~\ref{thm:main} after the analytic continuation.

\section*{Acknowledgments}
This work is supported in part by the Grant-in-Aid for Young
Scientists
from the Ministry of Education, Culture, Sports, Science and
Technology of Japan.


\end{document}